\documentclass[namedreferences]{kluwer}
\usepackage[]{graphicx}

\newcommand{\Ks}{\ensuremath{K_{S}}}
\newcommand{\JKs}{\ensuremath{J\!-\!K_{S}}}
\newcommand{\VKs}{\ensuremath{V\!-\!K_{S}}}
\newcommand{\bv}{\ensuremath{B\!-\!V}}
\newcommand{\etal}{et~al.}
\newcommand{\MvsdB}{\ensuremath{M_{V\!\mathrm{,sdB}}}}
\newcommand{\PopI}{Pop~\scalebox{0.85}{I}}
\newcommand{\chisq}{\ensuremath{\chi^{2}}}
\newcommand{\chisqr}{\ensuremath{\chi^{2}_{\mathrm{R}}}}

\begin{document}

\begin{article}
\begin{opening}         

\title{SINGLE AND COMPOSITE HOT SUBDWARF STARS IN THE LIGHT OF 2MASS
PHOTOMETRY} 

\author{M.~A. \surname{STARK}\email{stark@astro.psu.edu}}  
\author{RICHARD~A. \surname{WADE}\email{wade@astro.psu.edu}}  
\institute{Pennsylvania State University}

\author{G.~B. \surname{BERRIMAN}\email{gbb@ipac.caltech.edu}}  
\institute{IPAC, Caltech}

% If you really need this block in the article, then the figure size will
% need to be decreased to width=20pc
%\begin{ao}\\
%Department of Astronomy {\&} Astrophysics\\
%Pennsylvania State University\\
%525 Davey Lab\\
%University Park, PA, 16802, USA
%\end{ao}

\runningauthor{STARK, WADE, {\&} BERRIMAN}
\runningtitle{HOT SUBDWARFS IN 2MASS}

\date{June, 2003}

\begin{abstract}
Utilizing the Two Micron All Sky Survey (2MASS) All-Sky Data Release
Catalog, we have retrieved useful near-IR $J$, $H$, and {\Ks}
magnitudes for more than 800 hot subdwarfs (sdO and sdB stars) drawn
from the {\it Catalogue of Spectroscopically Identified Hot Subdwarfs}
(Kilkenny, Heber, {\&} Drilling 1988, 1992).  This sample size greatly
exceeds previous studies of hot subdwarfs.

We find that, of the hot subdwarfs in Kilkenny {\etal}, $\sim$40{\%}
in a magnitude limited sample have colors that are consistent with the
presence of an unresolved late type (FGK) companion.  Binary stars are
over-represented in a magnitude limited sample.  In an approximately
volume limited sample the fraction of composite-color binaries is
$\sim$25{\%}.
\end{abstract}


\begin{motto}[prose]
O star... Say something to us we can learn...
Say something! And it says, ``I burn.'' ...

\rightline{--Robert Frost}
\end{motto}
\end{opening}

\section{The Sample and Data Collection}  
We studied hot subdwarf stars listed in the {\it Catalogue of
Spectroscopically Identified Hot Subdwarfs} {\cite{KHD}} as updated in
an electronic version by Kilkenny, c.\,1992 (hereafter KHD).  Details
of the data collection and analysis procedure for a sub-sample from
the 2MASS $2^{\mathrm{nd}}$ Incremental Data Release can be found in
{\inlinecite{StarkWade}}.  In the current study we had improved
coordinates for {1486} objects (excluding 15 duplicates), while 26
objects remained unrecovered and are not included.  We recovered 2MASS
All-Sky counterparts for 1247 objects, 831 of which are hot subdwarfs
with {\it both} $J$ and {\Ks}.

For use in comparison, near-infrared colors for {\PopI} main sequence
(MS) stars {\cite{BB,Johnson}} were transformed to the 2MASS $JH{\Ks}$
system using relations given in the {\it Explanatory Supplement to the
2MASS All Sky Data
Release}\footnote{http://www.ipac.caltech.edu/2mass/releases/allsky/doc/explsup.html}.

\section{A ``Volume Limited'' Sample}  
Since 2MASS is a magnitude limited survey, 2MASS-KHD is a magnitude
limited sample (MLS).  In a MLS, {\it composite subdwarfs are
overrepresented} (the presence of a late type companion increases the
distance to which a composite remains brighter than the limiting
magnitude).  We made a cut in the ({\Ks}, {\JKs}) color-magnitude
diagram to define a {\it statistical} volume limited sample (VLS), by
removing composite subdwarfs that would not have been detected if they
were single.  {\inlinecite{StarkWade}} discuss this process in detail
(however, the cut made in the current study was updated and is thus
slightly different).

\section{Color Distribution}
We focused on {\JKs}, {\VKs}, and {\bv}.  Single hot subdwarfs mainly
are found in a region defined by: $\bv \lsim +0.1$, $\VKs \lsim +0.2$,
and $\JKs \lsim +0.05$.  Reported composite hot subdwarfs have redder
colors than single subdwarf stars (most notably in {\JKs}).

\begin{figure}
% larger figure size:
\centerline{\rotatebox{-90}{\includegraphics[width=22pc]{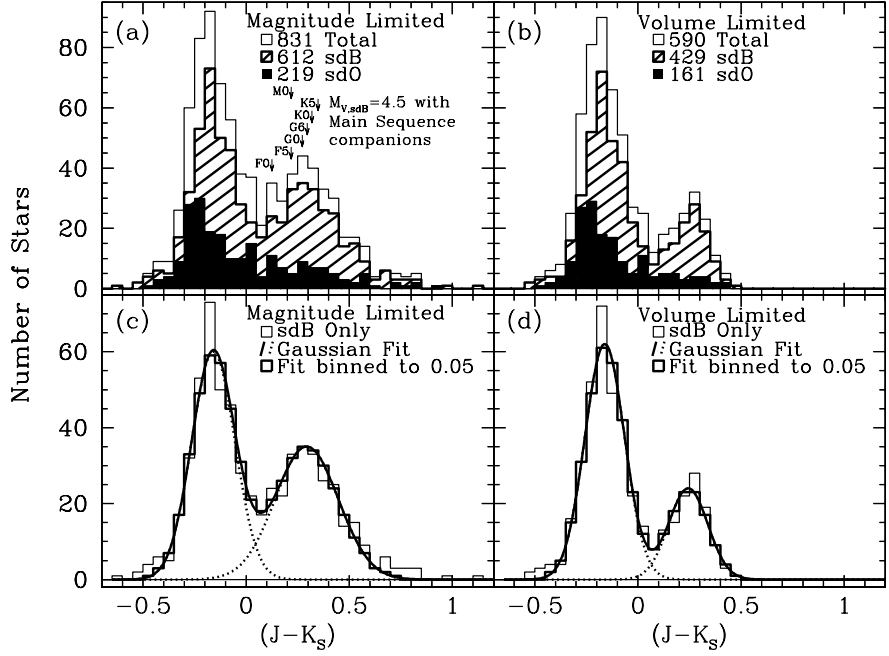}}}
% figure size if address block needs to be included at the end of paper:
%\centerline{\rotatebox{-90}{\includegraphics[width=20pc]{JKhistgauss.eps}}}
 \caption{Bimodal distribution of $\JKs$.  Panel $a$ also shows
 the composite $\JKs$ colors of sdB+MS stars assuming $M_V=4.5$
 and $\JKs=-0.15$ for the sdB.  Panels $c$ and $d$ show the Gaussian 
 fits to the sdB only distributions.}  \label{JKhist}
\end{figure}

Figure~{\ref{JKhist}} shows the distribution in {\JKs} for: the MLS
(panel $a$) and the VLS (panel $b$).
Both plots reveal a bimodal distribution (in both the total and sdB
only sub-sample) with peaks at: $\JKs \approx -0.15$ ($\sim$B3
spectral type) and $+0.30$ ($\sim$G0 spectral type).  Assuming stars
within the blue peak ($\JKs \lsim +0.05$) are single stars (or sd+WD
pairs, with colors indistinguishable from single hot subdwarfs), and
stars in the red peak ($\JKs \gsim +0.05$) are composite (sd+late
type) systems, then we obtain the binomial fractions of single and
composite-color systems observed by 2MASS in KHD given in
Table~\ref{composite}.
\begin{table}
 \caption{Binomial fractions of single and composite-color subdwarfs observed
 by 2MASS.} \label{composite}
\begin{tabular}{lcccccc} \hline
 {\bf Group}& & \multicolumn{2}{c}{{\bf MLS}}& & \multicolumn{2}{c}{{\bf VLS}} \\ 
            & & Single     & Composite  & & Single     & Composite  \\ \hline
 sdB        & & $55\pm4$\% & $45\pm4$\% & & $72\pm4$\% & $28\pm4$\% \\
 sdO        & & $67\pm6$\% & $33\pm6$\% & & $82\pm6$\% & $18\pm6$\% \\ \hline
 Total      & & $58\pm3$\% & $42\pm3$\% & & $75\pm3$\% & $25\pm3$\% \\ \hline
\end{tabular}
\end{table}

The bimodal distributions of sdB stars alone can be fit as the sum of
two Gaussians (parameters Table~\ref{Gaussian}, fits
Figure~\ref{JKhist}$c$--$d$).
\begin{table}
 \caption{Parameters for the Gaussian fits 
 (Figure~\ref{JKhist}$c$--$d$).} \label{Gaussian}
\begin{tabular}{lcccc} \hline 
                & \multicolumn{2}{c}{{\bf MLS}}  & \multicolumn{2}{c}{{\bf VLS}} \\
{\bf Parameter} & {\bf Blue} & {\bf Red} & {\bf Blue} & {\bf Red} \\ \hline
 Center         & $-0.159$   & $+0.294$  & $-0.161$   & $+0.244$  \\
 Amplitude      & $60$       & $35$      & $62$       & $24$      \\
 Dispersion     & $0.105$    & $0.153$   & $0.096$    & $0.095$   \\
Area Proportion & \multicolumn{2}{c}{54 : 46} & \multicolumn{2}{c}{72 : 28} \\ \hline
 Integral of Fit & \multicolumn{2}{c}{584} & \multicolumn{2}{c}{413} \\ 
{\chisq}, {\chisqr} (DOF) & \multicolumn{2}{c}{20.67, 1.03 (20)} & \multicolumn{2}{c}{11.96, 0.80 (15)} \\ \hline
\end{tabular}
\end{table}
We find that the Gaussian areas are consistent with the binomial
fractions from a cut at $\JKs = +0.05$.  The dispersions of both blue
Gaussians, and the VLS red Gaussian, are consistent with the average
2MASS photometric error [$\sigma(\JKs) \approx 0.1$], implying there
is little intrinsic spread in {\JKs} of single sdBs or the VLS late
type companions.

The strongly {\it bimodal} distribution in {\JKs} also indicates there
are no (or few) companions that are dM, or $\sim$F0 and earlier in
KHD.  In Figure~\ref{JKhist}$a$, the {\JKs} values of sdB+MS
composites are indicated assuming $\MvsdB = 4.5$ and
$(\JKs)_{\mathrm{sdB}} = -0.15$.  Companions of type F0 and earlier or
M0 and later would fall in the gap between the two peaks.  If a
significant population of such companions existed, the area between
the peaks would have been filled with their composites.

\section{Concluding Remarks}
The KHD catalog itself is likely {\it not} representative of the true
hot subdwarf population.  There are completeness issues, including
varying magnitude limits and classification criteria used by KHD's
sources.  Yet, since KHD is the most complete single compilation of
field hot subdwarfs available, we analyzed it despite its weaknesses.
Current and future surveys should result in a more complete census of
hot subdwarfs.

In KHD, we find $\sim$40{\%} MLS ($\sim$25{\%} VLS) of the subdwarfs
are consistent with having unresolved late type companions of spectral
types late F--K.  However, from these data alone we cannot determine
evolutionary states (MS and/or subgiant) unambiguously for these
companions, nor whether they interacted with the progenitor.

\acknowledgements
The authors thank E.\,M.\,Green for an annotated list of
spectroscopically observed hot subdwarfs, and R.\,F.\,Green for
helpful correspondence.  MS additionally thanks E.\,M.\,Green for the
exciting hike to the Keele observatory during the conference.  RW
thanks STScI for its hospitality.  This research has been supported
by: NASA grant NAG5-9586, NASA GSRP grant NGT5-50399, and a NASA Space
Grant Fellowship.  This research made use of: data products from the
Two Micron All-Sky Survey and the NASA/IPAC Infrared Science Archive.

\end{article}
\end{document}